\documentclass{aastex}
\usepackage{spr-astr-addons}
\RequirePackage{color}

\begin{document}
\title{Considering Late-Time Acceleration in some Cosmological Models}
\author{S. Davood Sadatian}
\affil{Department of Physics, Faculty of Basic Sciences University
 of Neyshabur, P. O. Box 91136-899, Neyshabur, Iran.\\
email : sd-sadatian@um.ac.ir}

\begin{abstract}
We study two cosmological models: A non-minimally coupled scalar
field on brane world model and a minimally coupled scalar field on
Lorentz invariance violation model. We compare some cosmological
results in these scenarios. Also, we consider some types of Rip
singularity solution in both models.

\end{abstract}

\section{Introduction}
In theories of extra spatial dimensions, ordinary matter is captured
on the brane but gravitation extends through the entire space-time
\citep{Ark98,Ran99,Dav00}. In these scenarios, the cosmological
evolution on the brane is taken by an effective Friedmann equation
that combines the effects of the bulk in a non-trivial
kind\citep{Bin00a,Bin00b,Maa04}. In other view point, the important
result in brane models is an alternative scenario for late-time
expansion of the universe. This result predicts deviations from the
4-dimensional gravity at limited distances. In other hand, the model
considered by Dvali, Gabadadze and Porrati (DGP) is different from
above models since it also predicts deviations from the standard
4-dimensional gravity in large distances\citep{Bin00a,Col00}.
Generally one can study the effect of a caused gravity term as a
quantum discipline in any brane world model. The existence of a
higher dimensional embedding space lets for the bulk or brane matter
that can evidently affect the cosmological evolution on the
brane\citep{Lan01}. A special form of bulk or brane context is a
scalar field. Scalar fields take an important key both in models of
the early universe and late-time acceleration. These scalar fields
give a dynamical model for matter fields in a brane world scenarios\citep{Bog06,Far06,Miz03,Bou05}.\\
In other hand, Lorentz invariance violation models (LIV) has been
considered in the scalar-vector-tensor theories\citep{Kan06}. It has
described that Lorentz violating vector fields influence the
dynamics equations in the inflationary models. An interesting result
of this model is that the exact Lorentz violating inflationary
solutions are depended on the absence of the inflation potential. In
this case, the inflation is exactly collaborated with the Lorentz
violation\citep{Ari07}. Therefore, we study this symmetry breaking
on the dynamics of equation of state for some cosmological aspects.\\
Some evidences from supernova data \citep{Per99,Rie98}, (CMB)
results \citep{Mil99,de00,Han00} and (WMAP) data
\citep{Spe03,Pag03,Spe07,Rei04,Zha12}, point out an accelerating
phase of cosmological expansion and this characterize shows that the
picture of universe by pressureless fluid is not enough; the
universe should contain some type of additional negative pressure
known as dark energy(For brief introduction in this field see
\citep{Ari07}). Also, the merged analysis of the WMAP data with the
supernova Legacy survey (SNLS)\citep{Spe07}, compels the equation of
state $w_{de}$, in accord with ${74\%}$ donation of dark energy in
the currently accelerating universe. Moreover, observations show
some sort of a dark energy equation of state, $w_{de}<-1$
\citep{Rei04}. Hence, a practical cosmological model should accept a
dynamical equation of state that may have crossed the value
$w_{de}= -1$, in late time of cosmological evolution.\\
In following, we consider cosmological results of a non-minimally
coupled scalar field on the brane and a minimally coupled scaler
field in LIV model. Also we determine late-time behavior of our
equations and obtain some restriction on the parameters
of models to have an accelerating universe.\\
We understand that dark energy is an important problem in modern
cosmology, especially, if the equation of state parameter $\omega$
less than $-1$. In this article, we study basically solving Friedman
equations and obtain the evolution of the effective equation of
state parameter $\omega$. The phantom crossing conditions that we
obtained for these set-ups, themselves are interesting at some
degree. Furthermore, we know that just considering background
quantity is not enough in such kind of unconventional cosmological
models. The current constraint on $\omega$ is obtained by combining
the result of the observation of CMB, which means that the
information of the evolution of perturbations should be also
included. Also we understand in the brane world model, the
perturbations of brane are coupled with these of bulk, which gives
nontrivial effect. Therefore, in first stage we just study the
background dynamics and the evolution of perturbations will study in future. \\
In other view, there is a no-go theorem which explicitly points out
a conventional dark energy model involving single degree of freedom
in the frame of standard Einstein gravity is forbidden to realize
such a scenario. This no-go theorem was proven in the appendix of
\citep{Xia08}. In this regards, we point out the original work in
\citep{Fen05} and several papers addressing this scenario
\citep{Cai07,Cai08,Cai10}. This theorem is exactly the reason why we
study a number of nonconventional dark energy models in realizing
the Quintom scenario\citep{Zha12}, such as the non-minimally coupled
on brane world model and the Lorentz-violating model considered in
the present paper. However, a {\it non-minimally} coupling scalar
field identified on the brane model and scalar field coupling {\it
minimally} to gravity in LIV model have some similar cosmological
results, this is an interesting theoretical motivation of these
models. Also we study other solutions admitted as Rip singularity,
that occur in the condition $\omega < -1$ increases rapidly.
However, it possible different types of singularity, depending
energy density and scale factor how increases with time
\citep{Fra11,Fra12}.

\section{Non-minimally coupled scalar Field on the Brane}

Here we study a brane world model where a scalar field is coupling
non-minimally to the Ricci scalar of the brane. In following we only
consider a scalar field in the matter Lagrangian without taking into
account baryons, cold dark matter, and radiation. This kind of
analysis suffers from a potential maybe cause that the model not be
able to explain the evolution of a realistic universe at background
level when confronting with observations. However, this problem
can solve by suitable fine-tuning parameters of model.\\
The action in the absence of ordinary matter can be given
as\citep{Bou05,Myu01}
\begin{equation}
S=\int d^{4}x\sqrt{-g}\bigg[\frac{1}{{k_{4}}^{2}}\alpha(\phi)
R[g]-\frac{1}{2} g^{\mu\nu} \nabla_{\mu}\phi\nabla_{\nu}\phi
-V(\phi) \bigg],
\end{equation}
where we have made a general non-minimal coupling $\alpha(\phi)$.
For simplicity, in following we set ${k_{4}}^{2}\equiv8\pi G_{N}=1$.
We can obtain Einstein equations with variation of the action
respect to brane metric
\begin{equation}
R_{\mu\nu}-\frac{1}{2}g_{\mu\nu}R=\alpha^{-1}{\cal{T}}_{\mu\nu}
\end{equation}
where ${\cal{T}}_{\mu\nu}$, energy-momentum tensor of the scalar
field non-minimally coupled to gravity is taken by
\begin{eqnarray}
{\cal{T}}_{\mu\nu}=\nabla_{\mu}\phi\nabla_{\nu}\phi-\frac{1}{2}g_{\mu\nu}(\nabla\phi)^{2}-g_{\mu\nu}V(\phi)
\nonumber
\end{eqnarray}
\begin{equation}
~~~~~+g_{\mu\nu}\Box\alpha(\phi)-\nabla_{\mu}\nabla_{\nu}\alpha(\phi),
\end{equation}
where $\Box$ shows 4-dimensional d'Alembertian. We study the FRW
universe with line element as following
\begin{equation}
ds^{2}=-dt^{2}+a^{2}(t)d{\Sigma_{k}}^{2}.
\end{equation}
where $d{\Sigma_{k}}^{2}$ is the line element for a constant
curvature $k = +1,0,-1$. The Ricci scalar obtain with the equation
of motion for scalar field as
\begin{equation}
\nabla^{\mu}\nabla_{\mu}\phi=V'-\alpha'R[g],
\end{equation}
where a prime denotes the derivative of each parameter with respect
to $\phi$, that may be written by
\begin{equation}
\ddot{\phi}+3\frac{\dot{a}}{a}\dot{\phi}+\frac{dV}{d\phi}=
\alpha'R[g].
\end{equation}
where a dot means the derivative of each parameter by respect to $t$
The intrinsic Ricci scalar for a FRW brane give as
\begin{equation}
R[g]=6\bigg(\dot{H}+2H^{2}+\frac{k}{a^{2}}\bigg),
\end{equation}
and Friedmann's equations are determined by
\begin{equation}
\frac{\dot{a}^{2}}{a^{2}}=-\frac{k}{a^{2}}+\frac{\rho}{3},
\end{equation}
and
\begin{equation}
\frac{\ddot{a}}{a}=-\frac{1}{6}(\rho+3p).
\end{equation}
We take a scalar field, $\phi$, only depended on time. So, with
Eq(3), we obtain
\begin{equation}
\rho=\alpha^{-1}\bigg(\frac{1}{2}\dot{\phi}^{2}+V(\phi)-6\alpha'H\dot{\phi}\bigg),
\end{equation}
\begin{equation}
p=\alpha^{-1}\bigg(\frac{1}{2}\dot{\phi}^{2}-V(\phi)+
2\Big(\alpha'\ddot{\phi}+2H\alpha'\dot{\phi}+\alpha''\dot{\phi}^2\Big)\bigg)
\end{equation}
where $H=\frac{\dot{a}}{a}$ is Hubble parameter. Now equation of
state has the following form
\begin{equation}
w\equiv\frac{p}{\rho}=\frac{\dot{\phi}^{2}-2V(\phi)+
4\Big(\alpha'\ddot{\phi}+2H\alpha'\dot{\phi}+\alpha''\dot{\phi}^2\Big)}{\dot{\phi}^{2}+2V(\phi)-12\alpha'H\dot{\phi}}.
\end{equation}
When $\dot{\phi}=0$, we have $p=-\rho$. In this illustration $\rho$
depended on $a$ and $V(\phi)$, it has the duty of a cosmological
constant. In the minimal case which $\dot{\phi}^{2}< V(\phi)$, by
Eq(9), we take $p<-\frac{\rho}{3}$ that shows a late-time
accelerating universe. In non-minimal case, the position depends on
the choice of non-minimal coupling parameter. In following we show
how for a suitable range of coupling  parameter, a late-time
accelerating expansion can be described. We first consider a moving
domain wall of brane world to discuss quintessence behavior, then we
study a special non-minimal coupling for late-time acceleration.
\subsection{Late-Time Acceleration in a Brane world Model}
We study a bulk config by two 5-dimensional anti de
Sitter-Schwarzschild (AdSS$_{5}$) black hole spaces combined on a
{\it moving domain wall}. To insert this moving domain wall into
5-dimensional bulk, it is required to indicate normal and tangent to
the domain wall by determination of normal instructing to the brane.
We take that domain wall is identified at coordinate $r=a(\tau)$
where $a(\tau)$ is considered by Israel junction conditions
\citep{Car04}. Here we study the following line element\citep{Myu01}
\begin{eqnarray}
{{ds}_{5\pm}}^{2}=-\bigg(k-\frac{\eta_{\pm}}{r^2}+\frac{r^2}{\ell^{2}}\bigg)dt^{2}\nonumber
\end{eqnarray}
\begin{equation}
~~~~+\frac{1}{k-\frac{\eta_{\pm}}{r^2}+\frac{r^2}{\ell^{2}}}dr^{2}+r^{2}\gamma_{ij}dx^{i}dx^{j},
\end{equation}
where $\pm$ is for left($-$) and right($+$) side of the moving
domain wall, also $\ell$ is curvature radius of AdS$_{5}$ manifold
and $\gamma_{ij}$ is the horizon metric. $\eta_{\pm}\neq 0$ creates
the electric part of the Weyl tensor on two sides\citep{Myu01}.\\
We assume usual mater on the brane has a perfect fluid form,
$T_{\mu\nu}=(\rho+p)u_{\mu}u_{\nu}+p h_{\mu\nu}$ where
$\rho=\rho_{m}+\sigma$ and $p=p_{m}-\sigma$. Energy density of the
confined matter on the brane and analogous pressure are shown with
$\rho_{m}$ and $p_{m}$, while $\sigma$ is the brane tension. with
Israel junction conditions\citep{Car04} and Gauss-Codazzi equations,
we obtain generalized Friedmann equations\citep{Myu01} as following
\begin{equation}
\frac{\dot{a}^{2}}{a^2}+\frac{k}{a^{2}}=\frac{\rho_{m}}{3}+\frac{\eta}{a^{4}}+
\frac{\ell^{2}}{36}\rho_{m}^{2}
\end{equation}
\begin{equation}
\frac{\ddot{a}}{a}=-\frac{\rho_{m}}{6}(1+3w)-\frac{\eta}{a^4}-\frac{\ell^2}{36}\rho_{m}^{2}(2+3w)
\end{equation}
where we require a $Z_{2}$-symmetry by $\eta_{+}=\eta_{-}\equiv\eta$
and we have supposed $p_{m}=w\rho_{m}$. following we take case:
$\eta=0$. For $\eta=0$, each sub-manifolds of bulk space-time are
accurate AdS$_{5}$ space-times. Now we study a confined
non-minimally coupled scalar field on the brane and consider
cosmological aspects. We use energy density and pressure of scalar
field (10) and (11) and by equation (15), hence for cosmic expansion
have
\begin{eqnarray}
\frac{\ddot{a}}{a}=-\frac{1}{6\alpha}\bigg(\frac{1}{2}\dot{\phi}^{2}+V(\phi)-6\alpha'H\dot{\phi}\bigg)\times
\nonumber
\end{eqnarray}
\begin{eqnarray}
\bigg(1+3\frac{\dot{\phi}^{2}-2V(\phi)+
4\big(\alpha'\ddot{\phi}+2H\alpha'\dot{\phi}+\alpha''\dot{\phi}^2\big)}{\dot{\phi}^{2}+2V(\phi)-12\alpha'H\dot{\phi}}
\bigg) \nonumber
\end{eqnarray}
\begin{eqnarray}
-\frac{\ell^2}{36\alpha^2}\bigg(\frac{1}{2}\dot{\phi}^{2}+V(\phi)-6\alpha'H\dot{\phi}\bigg)^{2}\times
\nonumber
\end{eqnarray}
\begin{equation} \bigg(2+3\frac{\dot{\phi}^{2}-2V(\phi)+
4\big(\alpha'\ddot{\phi}+2H\alpha'\dot{\phi}+\alpha''\dot{\phi}^2\big)}{\dot{\phi}^{2}+2V(\phi)-12\alpha'H\dot{\phi}}\bigg).
\end{equation}
where $H=\frac{\dot{a}}{a}$ is Hubble parameter. This is a complex
equation, to describe cosmological aspects, we must study some
limiting cases or specify $\alpha(\phi)$, $V(\phi)$ and $\phi$.\\
Equation (16) shows, the condition for an accelerating universe
depended on the choice of non-minimal coupling and the scalar field
potential. Hence we determine the following non-minimal coupling
\begin{equation}
\alpha(\phi)=\frac{1}{2}\Big(1-\xi \phi^{2}\Big).
\end{equation}
For an accelerating universe we should have $\rho_{m}+3p_{m}<0$
\citep{Bou05,Far00}. If we take $V>0$, we obtain
\begin{equation}
(1-3\xi)\dot{\phi}^{2}-V(\phi)-3\xi\phi(\ddot{\phi}+H\dot{\phi})<0
\end{equation}
dynamics of this scalar field is given as
\begin{equation}
\ddot{\phi}+3H\dot{\phi}+\xi R\phi+\frac{dV}{d\phi}=0,
\end{equation}
we obtain
\begin{equation}
(1-3\xi)\dot{\phi}^{2}-V(\phi)+3\xi^{2}R\phi^{2}+6\xi
H\phi\dot{\phi}+3\xi\phi\frac{dV}{d\phi}<0.
\end{equation}
With $\rho_{m}$ in this case, we determine
\begin{eqnarray}
y\equiv(1-\xi\phi^{2})\rho_{m}-2V(\phi) \nonumber
\end{eqnarray}
\begin{equation}
+\dot{\phi}^{2}\Big(\frac{1}{2}-3\xi\Big)+3\xi^{2}R\phi^{2}+3\xi\phi\frac{dV}{d\phi}<0
\end{equation}
that is the requirement to have an accelerated universe. In
following, we study the weak energy condition $\rho_{m}\geq0$ and
limit our consideration to the case by $\xi\leq 1/6$. Therefore we
obtain
\begin{equation}
-2V+3\xi\phi\frac{dV}{d\phi}<y<0
\end{equation}
and a main condition for cosmic acceleration is
\begin{equation}
V-\frac{3\xi}{2}\phi\frac{dV}{d\phi}>0,\quad\quad\quad
\xi\leq\frac{1}{6}.
\end{equation}
\begin{figure}[h]
\begin{center}\includegraphics{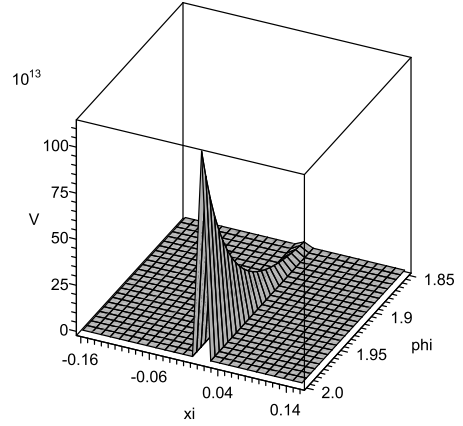} \vspace{6.5cm}
\end{center}
 \caption{\small {$V$ vary
for different values of the $\phi$ and Non-minimal coupling
parameter $\xi$ for $V = \phi^{\frac{2}{3\xi}}$.}}
\end{figure}
For the model have cosmic acceleration by $\xi>0$, the potential
$V(\phi)$ must change with $\phi$\, slower than power-law potential
$V_{c}(\phi)=V_{0}\Big(\frac{\phi}{\phi_{0}}\Big)^{\frac{2}{3\xi}}$
(Fig 1). In other hand, when $\xi<0$, the main condition for cosmic
acceleration needs $V$ grow faster than $V_{c}$ as $\phi$
increases\citep{Far00}. Now we take a simple special example to show
how this model process. By potential of the form $V(\phi)=\lambda
\phi^{n}$,\, condition (23) takes
$\lambda\Big(1-\frac{3n\xi}{2}\Big)>0$\,. In this case just
for $\xi\leq 2/3n$ we have the accelerated expansion.\\

More general case is $\eta\neq 0$ in equations (14) and (15). In
following we convert the generalized Friedmann equation as
\begin{eqnarray}
\dot{\chi}+4\frac{\dot{a}}{a}\bigg[\chi+\frac{1}{12\alpha}\bigg(\frac{1}{2}(1-2\xi)\dot{\phi}^{2}-V(\phi)
-2\xi\phi\Big(\ddot{\phi}+3\frac{\dot
a}{a}\dot{\phi}\Big)\bigg)\bigg]\nonumber
\end{eqnarray}
\begin{equation}
=0,
\end{equation}
where $\chi=\Big(\frac{\dot a}{a}\Big)^{2}$ is dark energy field.
Then we take $\phi(t)\approx\frac{A}{t^{\beta}}$\, where $A$ is a
constant and $\beta$ an unspecified power will be studied. For scale
factor we take the ansatz $a(t)\approx Bt^{\nu}$ where $B$ is a
constant and $\nu$ to be determined. For scalar field potential we
take $V(\phi)=\lambda\frac{\phi^{2}}{2}$. Therefore we obtain
\begin{eqnarray}
\frac{\ddot{a}}{a}+\Big(\frac{\dot a}{a}\Big)^{2}\approx
\frac{\sigma^{2}}{12\alpha u}+ \frac{\sigma}{24\alpha
u}\rho(1-3w)+\frac{1}{6\alpha}\Big(V(\phi)+\Lambda\Big) \nonumber
\end{eqnarray}
\begin{equation}
~~~~-\frac{\dot{\phi}^{2}}{12\alpha}(1+4\alpha'')-
\frac{\alpha'}{3\alpha}\Big(\ddot{\phi}+3\frac{\dot
a}{a}\dot{\phi}\Big)
\end{equation}
and
\begin{eqnarray}
\ddot{\phi}+3\frac{\dot{a}}{a}\dot{\phi}+\frac{u'}{2u}\dot{\phi}^{2}\approx
-\frac{6\alpha}{u}V'+10\frac{\alpha'}{u}\Big(V(\phi)+\Lambda\Big)
\nonumber
\end{eqnarray}
\begin{equation}
+4\frac{\alpha'(u+\alpha'
u')\sigma}{u^{3}}\rho(1-3w)+\frac{8\alpha'(u+\alpha'
u')}{u^3}\sigma^{2}
\end{equation}
where $u=6\Big[1-\xi(1-\frac{16}{3}\xi)\frac{\phi^{2}}{2}\Big]$. We
take usual matter on the brane has an equation of state such as\,
$\rho=Dt^{-3\nu(1+w)}$. Hence we have
\begin{eqnarray}
\frac{\nu(2\nu-1)}{t^{2}}\approx\frac{1}{24}\Bigg\{\frac{\sigma^2}{3}+4\Lambda+
\bigg(\frac{\sigma(1-3w)D}{6}\bigg)\frac{1}{t^{3\nu(1+w)}}+
\nonumber
\end{eqnarray}
\begin{eqnarray}
\frac{A^2}{t^{2\beta}}\bigg[2\lambda+2\xi\bigg(\Lambda+\frac{4(\frac{3}{8}-\xi)}{9}\sigma^2\bigg)+
\frac{4\xi D(1-3w)(\frac{3}{8}-\xi)\sigma}{9}\times \nonumber
\end{eqnarray}
\begin{equation} \frac{1}{t^{3\nu(1+w)}}
\frac{2\beta}{t^2}\bigg(\beta(1-8\xi)+4\xi(3\nu-1)\bigg)\bigg]\Bigg\}
\end{equation}
and scalar field equation given as
\begin{eqnarray}
\beta\Big(\beta+1-3\nu\Big)\frac{A}{t^{\beta+2}}\approx-\frac{\lambda
A}{t^{\beta}} \bigg[1-\frac{8}{3}\xi^2\frac{A^2}{t^{2\beta}}\bigg]
\nonumber
\end{eqnarray}
\begin{eqnarray}
-\frac{5\xi}{3}\frac{A}{t^{\beta}}\bigg[\Lambda+\Big(\frac{\lambda}{2}+\frac{\xi}{2}\big(1-\frac{16}{3}\xi\big)\Lambda\Big)\frac
{A^2}{t^{2\beta}}\bigg] \nonumber
\end{eqnarray}
\begin{equation}
-\frac{\xi
\sigma}{9}\bigg(2\sigma+\frac{D}{t^2}(1-3w)\bigg)\frac{A}{t^{\beta}}\bigg[1+\xi(1-\frac{16}{3}\xi)(1+\xi)\frac{A^2}{t^{2\beta}}\bigg]
\end{equation}
where $\beta >1$. Two conditions must solve numerically to limit the
values of $\nu$ and $\beta$ for an accelerating expansion. After
some calculations we obtain
\begin{equation}
-\frac{(1-3w)\xi\sigma}{9}D=A\beta(\beta+1-3\nu),
\end{equation}
\begin{equation}
\lambda=\frac{-\xi\sigma^2}{12},
\end{equation}
where we passed over terms of order ${\cal O}(t^{-3\beta})$ and
higher. Now, if we arrange $\lambda$, since
$\nu(2\nu-1)=\frac{\sigma(1-3w)}{144}D$, \,we find
\begin{equation}
\beta=\frac{3\nu-1}{2}\pm\bigg[\frac{(3\nu-1)^2}{2}-16\xi\nu(2\nu-1)\bigg]^{\frac{1}{2}},
\end{equation}
where $\nu=\frac{2}{3(1+w)}$. For $\nu<1$ the parameter $\beta$ is
real for any value of $\xi$ , for $\nu>1$ the exponent $\beta$ is
real only if we have
\begin{equation}
\xi\leq\frac{1}{16}\frac{(3\nu-1)^{2}}{2\nu(2\nu-1)}.
\end{equation}
\begin{figure}[h]
\begin{center}\includegraphics{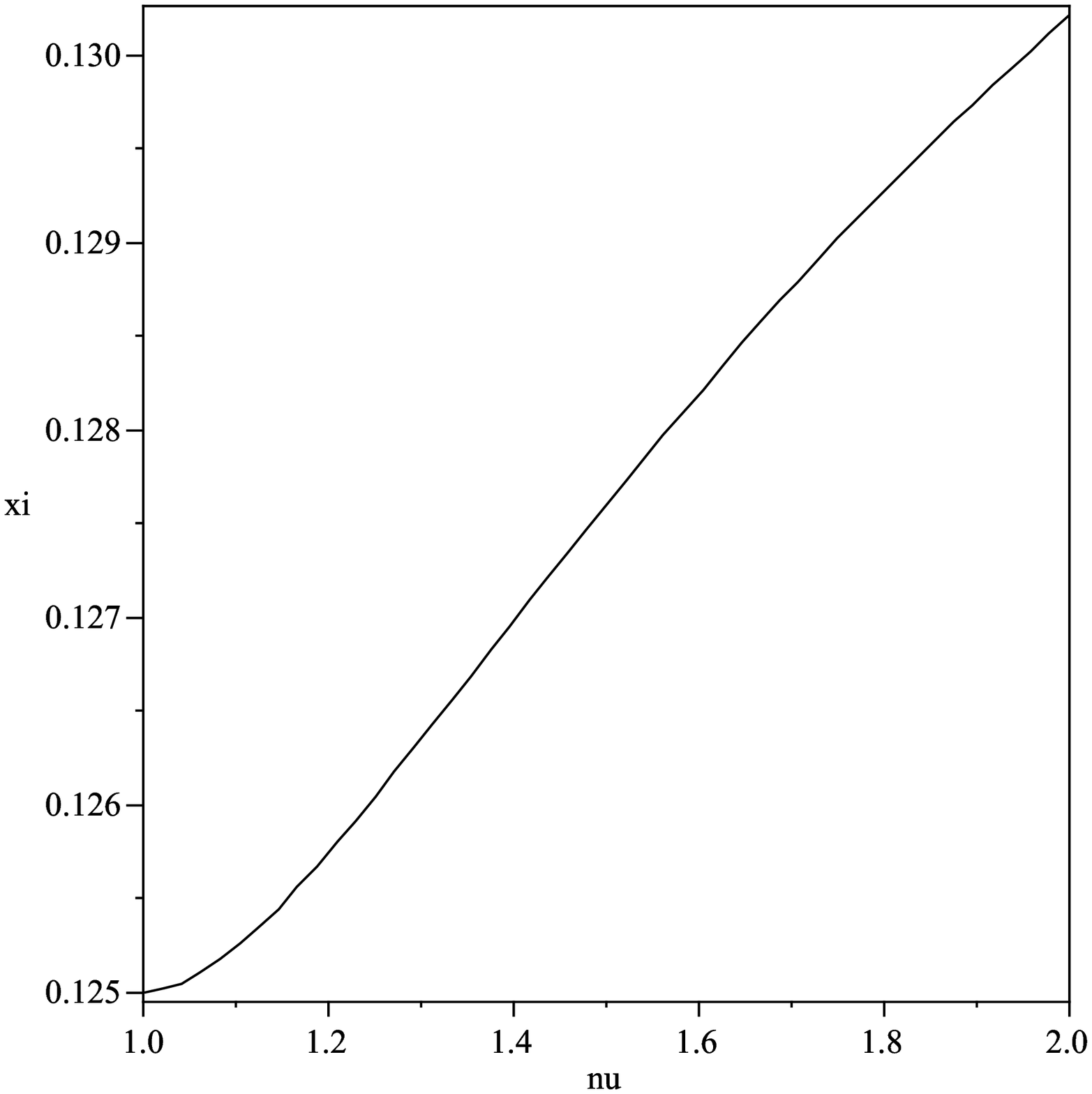} \vspace{11cm}\includegraphics{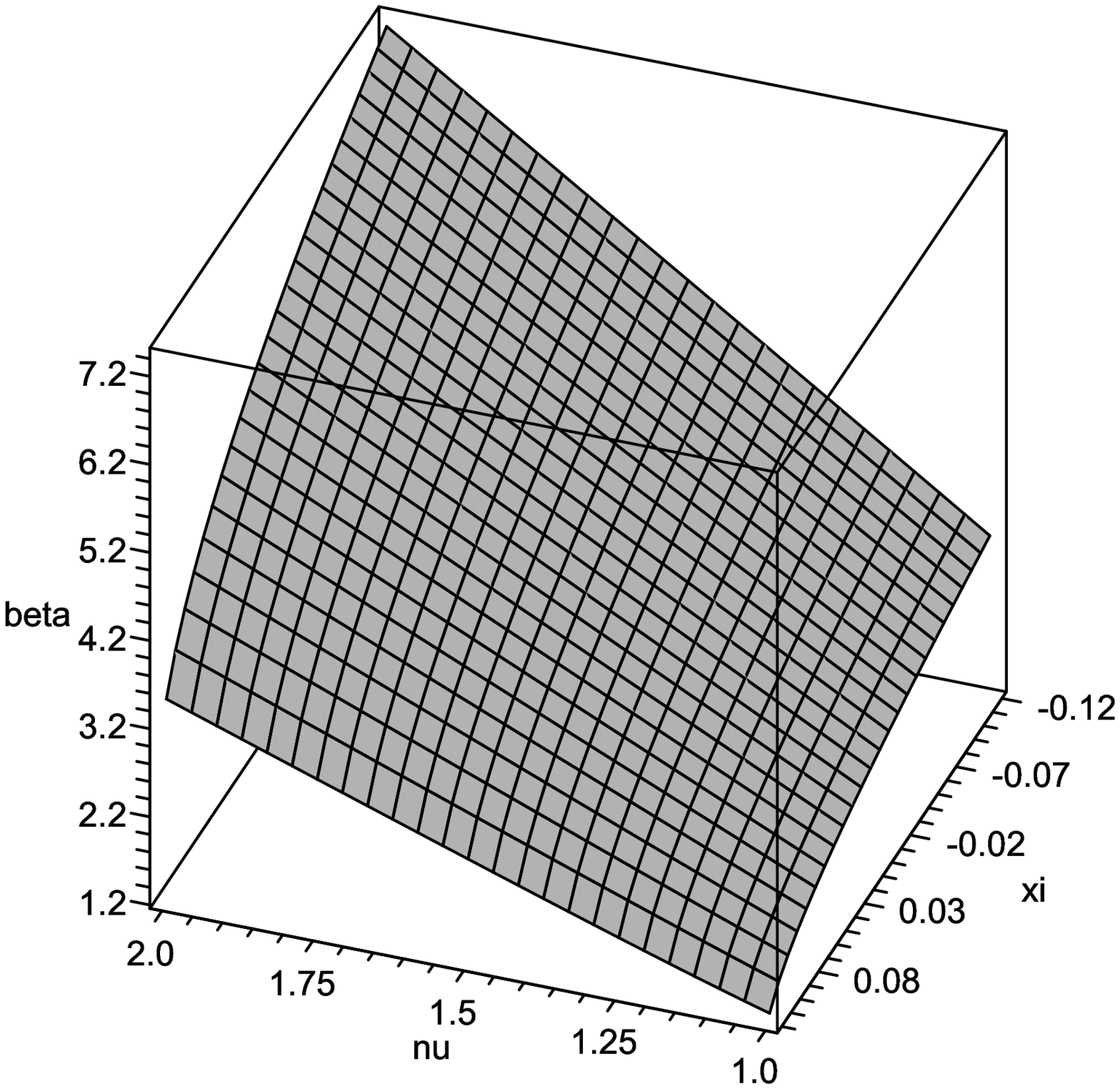}
\end{center}
\caption{\small {Variation of $\xi$ respect to $\nu$ with equation
(32) (up) and Variation of $\beta$ respect to $\nu$ and $\xi$ with
equation (31) by sign $+$, for sing $-$ in this equation (31)
$\beta$ has negative value. We choose special range of parameter
$\xi$ with equation (32) in (down) figure. }}
\end{figure}
These equations limited the values of non-minimal coupling $\xi$ for
obtain an accelerating expansion(Fig 2). Hence, it seems that
non-minimally coupled scalar field identified on the
brane world supplies natural candidate for late-time expansion.\\
Also we can use equation (12) for determine dynamical Equation of
state $\omega(t)$.
\begin{figure}[h]
\begin{center}\includegraphics{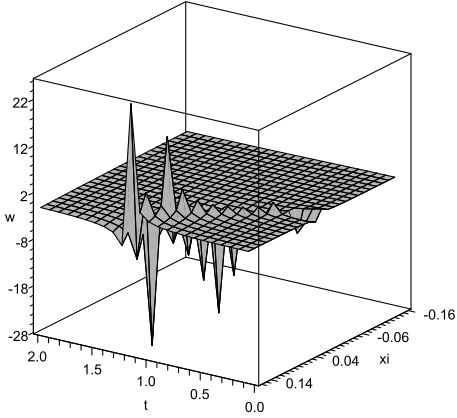} \vspace{7cm}
\end{center}
 \caption{\small {Dynamical Equation of state $\omega(t)$
vary for different values of the $t$ and Non-minimal coupling
parameter $\xi$  by equation (12) and specific choose for parameters
space.}}
\end{figure}

Figure (3) shows possible crossing of phantom divided barrier in
this framework, this result is important because previous
considerations have shown, crossing phantom divided line with a
scalar field non-minimally coupling gravity exist\citep{Nes07,Vik05}.\\
In following, we explain the physical reasons why the crossing of
the phantom divide can occur in this model. Here, we have a scalar
field that non minimally coupled in gravity and an ordinary matter.
Generally, according to \citep{Nes07}, these sources of
energy-momentum can switch together and describe evolution of
universe. In this regard, an encouraging candidate for the dark
matter is either a positive cosmological constant or a slowly
developing scalar known as "quintessence". Also the quintessence can
be determined as an alternative method to solve the cosmological
constant problem. This is possible because instead of the
fine-tuning, the quintessence gives a model of slowly decaying
cosmological constant. However, there exists another question for
the quintessence now. This is to inquire whether the expansion will
keep accelerating forever or it will slow down  again after some
time. This is very similar to the leaving problem in the
inflationary cosmology. According to the theory of quintessence, the
dark energy of the universe is controlled by the potential of a
scalar field $\phi$ which is still curling to its minimum at V = 0.
\subsection{Rip Singularity}
The new interest to the models with the crossing phantom divide
$\omega< -1$ is their prediction of a Rip singularity \citep{Sta00}.
Generally, the scale factor of the universe gets infinite at a
finite time in the future which was named Big Rip singularity. There
were proposed some scenarios to improved the Big Rip singularity:
(I) To determined phantom acceleration as temporary phenomenon. This
is a number of scalar potentials. (II) To explain for quantum
effects which maybe delay/stop the singularity happening
\citep{Eli04}. (III) To change the gravitation itself in such a way
that it seems to be observationally-friendly from a side but it
restores to  singularity. (IV) To couple dark energy with dark
matter in the specific way or to use specific form for dark energy
equation of state \citep{Bam08}. Pay attention to for quintessence
dark energy, other finite-time singularities may appear. For
instance, type II (sudden) singularity or type III singularity
appears with finite scale factor but infinite energy and/or
pressure. The near examination shows that the situation $\omega <
-1$ is not enough for a singularity event\citep{Noj05}. First of
all, a transient phantom cosmology is possible. However, one can
make such models that $\omega$ asymptotically tends to -1 and the
energy density increases with time or corpses constant but there is
no finite-time singularity \citep{Noj05}. Of course, most apparent
case is when Hubble rate inclined to constant (cosmological constant
or asymptotically de Sitter space), which can also point out the
pseudo-rip \citep{Fra12}. most interesting situation is connected
with Little Rip cosmology \citep{Fra11} where Hubble rate go to
infinity in the infinite future. The goal point is that if $\omega$
approaches -1 adequately rapidly, then it is possible to have a
model in which the time needed for
singularity is infinite. However, this situation does Not occur in our model, because $H$ is finite. \\
In following, if we want to determind some types of Rip singularity,
we must use Hubble parameter, energy density and
pressure equations that obtained in above equations. Therefore, after calculations with $\xi=0.14$ (figure 4), we take results as following:\\

$case\, 1$ : type II singularity (sudden)\\
conditions: scale factor and energy density arrives finite value but
$p\rightarrow \infty$

In this model, we find that scale factor, pressure and energy
density are finite. Therefore, the model does not predict a sudden
singularity.

$case\, 2$ : type III singularity\\
Conditions, scale factor arrives finite value but ($\rho\rightarrow
\infty$) and $\mid p\mid\rightarrow \infty$

This means the energy density becomes so rapidly with time and scale
factor does not arrive the infinite value. The goal difference
between case (1) and (2) for this model is that the potential of
scalar field has a stake in a case of singularity. Here, according
numerical calculations, if scale factor arrives finite value, energy
density and pressure are finite. Hence, type III singularity does
not occur in this setup.

$case\, 3$ : pseudo-rip singularity\\
conditions : Pressure $p\rightarrow-\rho$ in infinite time and cause
to $\dot{H}\rightarrow 0$

This means the expansion of the universe asymptotically appeals to
the exponential area. Obviously in our model with above setup,
pseudo-rip singularity predict, because in infinite time, when
$\dot{H}$ arrive to zero, $p\rightarrow-\rho$.
\begin{figure}[h]
\begin{center}\includegraphics{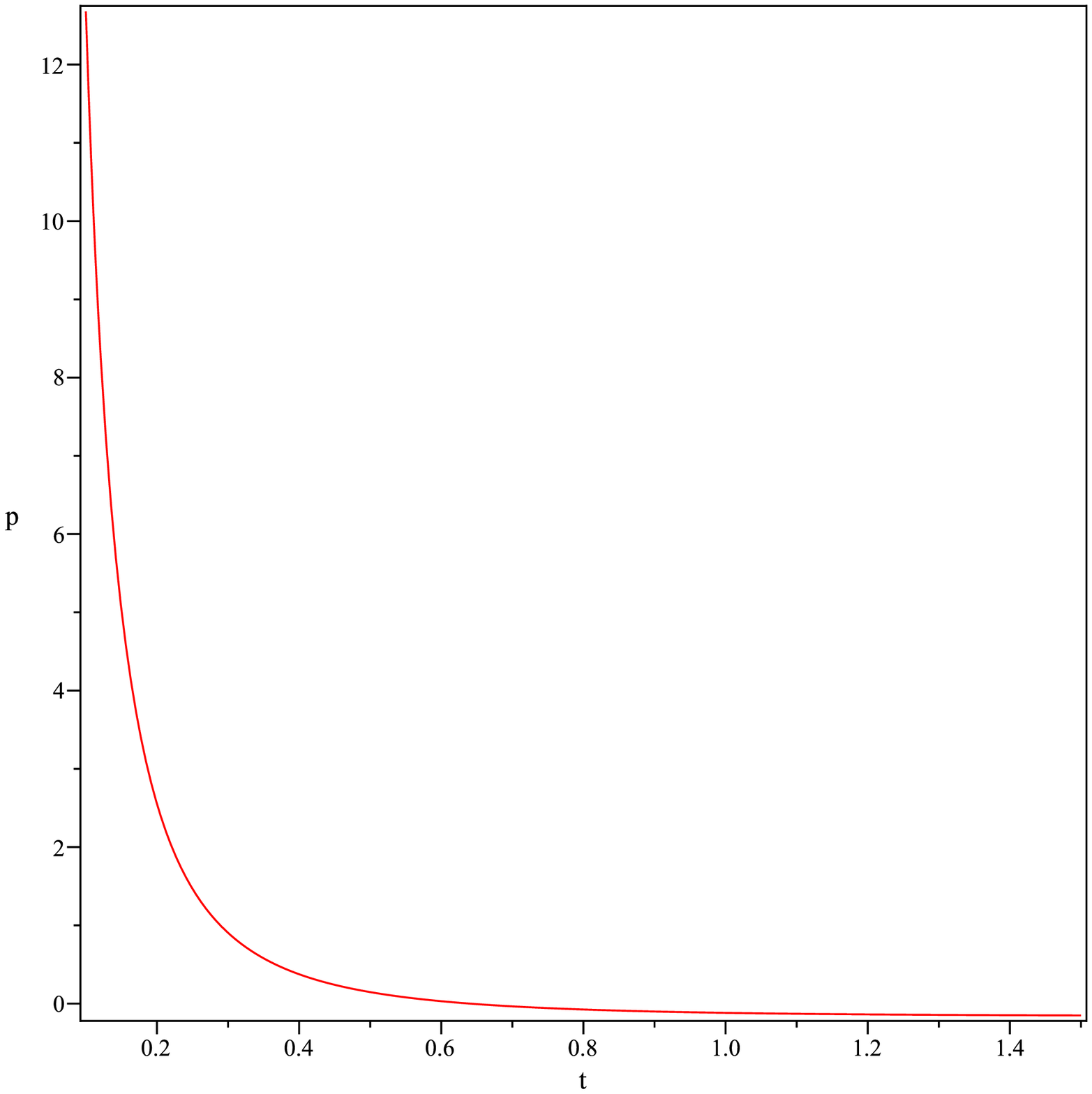} \vspace{11cm}\includegraphics{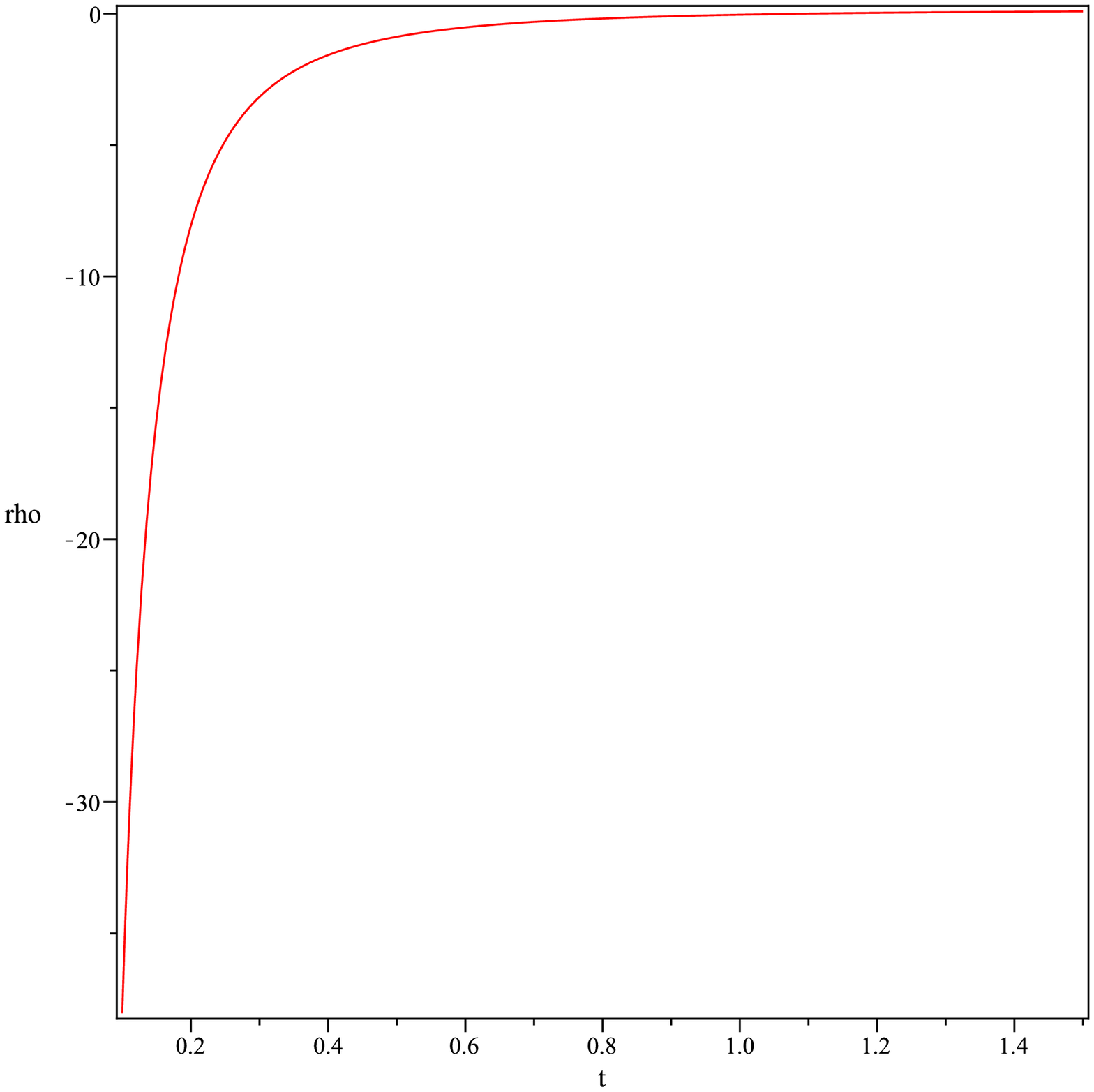}
\end{center}
\caption{\small {Variation of pressure(up) and energy density(down)
respect to time for a non minimal coupling model(where $\xi=0.14$)
}}
\end{figure}

\section{A Lorentz Violating Model}

With using \citep{Kan06,Ari07,sad12}, we consider the cosmological
aspects of Lorentz invariance violating model. We emphasize this
model parameters are tightly bounded by both astronomical and
cosmological experiments, some works such as
\citep{Ari07,Ari08,Fre09,Ari10} explained, the model parameters are
able to be bounded by comparing the model with cosmological
observations. Also a global analysis and a group of canonical values
for the model parameters to constrain these parameters given in
\citep{Kos86}, so that the model is expected to be viable and
background
evolution is reasonable.\\
We want to obtain a connection between Lorentz Invariance violation
parameter and dynamics of scalar field. First we define an action
for a representative scalar-vector-tensor theory which allows
Lorentz invariance violation as
\begin{eqnarray}
     S&=& S_g + S_u + S_{\phi} \ ,
     \end{eqnarray}
where the actions for the vector field $S_u$, the tensor field
$S_g$, and the scalar field $S_{\phi}$ are assume as
\begin{eqnarray}
S_g=\int d^4 x \sqrt{-g}~ {1\over 16\pi G}R
\end{eqnarray}
\begin{eqnarray}
S_u=\int d^4 x \sqrt{-g}[ - \beta_1 \nabla^\mu u^\nu \nabla_\mu
u_\nu-\beta_2 \nabla^\mu u^\nu \nabla_\nu u_\mu \nonumber
\end{eqnarray}
\begin{equation}
-\beta_3 \left( \nabla_\mu u^\mu \right)^2-\beta_4 u^\mu u^\nu
\nabla_\mu u^\alpha \nabla_\nu u_\alpha+ \lambda \left( u^\mu u_\mu
+1 \right)]
\end{equation}
\begin{eqnarray}
S_{\phi}=\int d^4 x \sqrt{-g}~ {\cal{L}}_{\phi}.
\end{eqnarray}
This action has a non-gravitational degrees of freedom in the
structure of Lorentz violating scalar-tensor-vector theory. We take
$u^\mu u_\mu = -1$ and the expected value of vector field $u^\mu$ is
$<0| u^\mu u_\mu |0> = -1$\,\citep{Kos86}. $\beta_i(\phi)$
($i=1,2,3,4$) are unrestricted parameters by dimension of mass
squared , also ${\cal{L}}_{\phi}$ is the Lagrangian density for the
scalar field in this model\citep{Kan06}. If universe be homogeneous
and isotropic, we consider the universe by metric as
\begin{eqnarray}
ds^2 = - {\mathcal{N}}^2 (t) dt^2 + e^{2\alpha(t)} \delta_{ij} dx^i
dx^j \ ,
\end{eqnarray}
where ${\mathcal{N}}$ is a lapse function and the universe scale is
given by $\alpha$\,\citep{Kan06,Ari07}.  If the action vary with
respect to metric and selecting a fitting gauge, field equations
obtain as
\begin{eqnarray}
   R_{\mu\nu}-{1\over 2}g_{\mu\nu}R = 8\pi G T_{\mu\nu} \ ,
   \end{eqnarray}
where $T_{\mu\nu} =T_{\mu\nu}^{(u)} + T_{\mu\nu}^{(\phi)}$ is the
total energy-momentum tensor, and energy-momentum tensors of vector
and scalar fields given by $T_{\mu\nu}^{(\phi)}$. The time and space
elements of the total energy-momentum tensor obtain as \citep{Ari07}
\begin{eqnarray}
     T^{0}_{0} = - \rho_u -\rho_{\phi} \ , \qquad    T^{i}_i =  p_u+ p_{\phi} \ ,
     \end{eqnarray}
where the energy density and pressure of the vector field determine
as
\begin{eqnarray}
    && \rho_u =  -3\beta H^2  \ ,
   \\
    &&p_u =  \left(3 + 2{H^{\prime}\over H} + 2{\beta^{\prime}\over \beta} \right)\beta H^2 \ ,
     \\
    && \beta \equiv \beta_1 +3 \beta_2 + \beta_3 \ ,
    \end{eqnarray}
a prime point to the derivative of any parameters by respect to
$\alpha$ \, and \, $H\equiv d\alpha/dt=\dot{\alpha}$ is the Hubble
parameter. The energy equations for scalar field, $\phi$ and the
vector field $u$ given as
\begin{eqnarray}
   {\rho}^{\prime}_u + 3({\rho}_u + p_u)=+3H^2 \beta^{\prime}  \ ,
   \end{eqnarray}
\begin{eqnarray}
    {\rho}^{\prime}_{\phi} + 3({\rho}_{\phi} + p_{\phi})=-3H^2 \beta^{\prime}  \
    ,
   \end{eqnarray}
So, the total energy equation for two the vector and the scalar
fields write as
\begin{equation}
   {\rho}^{\prime} + 3({\rho} + p)=0 \ , \quad (\rho = \rho_u + \rho_{\phi}) .
   \end{equation}
Now, dynamics of the model determine with the following Friedmann
equations\citep{Kan06,Ari07}
\begin{eqnarray}
\left( 1 + \frac{1}{8\pi G \beta} \right) H^2={1\over 3\beta}
\rho_{\phi}
\end{eqnarray}
\begin{eqnarray}
(1 + \frac{1}{8\pi G \beta} )( HH'+H^2)=-{1\over 6}(
{\rho_{\phi}\over \beta} + {3p_{\phi}\over \beta}) - H^2
{\beta'\over \beta}.
\end{eqnarray}
For the scalar section of this model we assume the
following Lagrangian
\begin{eqnarray}
{\cal{L}}_{\phi}= -{\eta\over 2}(\nabla \phi)^2 - V(\phi) \ ,
\end{eqnarray}
where $(\nabla
\phi)^2=g^{\mu\nu}\partial_{\mu}\phi\partial_{\nu}\phi$. Usual
scalar fields are match to $\eta = 1$ while $\eta = -1$ describe
phantom fields. The homogeneous scalar field has the density
$\rho_{\phi}$ and pressure $p_{\phi}$ as
\begin{eqnarray}
&&\rho_{\phi} = {\eta\over 2} H^2 \phi^{\prime 2} + V(\phi) \ ,
\\
&&p_{\phi}= {\eta\over 2} H^2 \phi^{\prime 2} - V(\phi) \
\end{eqnarray}
and equation of state parameter obtain as
\begin{eqnarray}
\omega_{\phi}={p_{\phi}\over \rho_{\phi}} = - \frac{1- \eta H^2
\phi^{\prime 2}/2V}{1 + \eta H^2 \phi^{\prime 2}/2V} \ .
\end{eqnarray}
Now the Friedmann equation take as\,\citep{Ari07}
\begin{eqnarray}
H^2  = \frac{1}{3\bar{\beta}} \left[
  \frac{\eta}{2} H^2 \phi^{\prime 2} + V(\phi) \right] ,
  \end{eqnarray}
where $\bar{\beta}=\beta+\frac{1}{8\pi G}$. With this equation we
obtain
\begin{eqnarray}
  \phi^{\prime} &=&-2\eta \bar{\beta}\left(\frac{H_{,\phi}}{H} +
\frac{\bar{\beta}_{,\phi}}{\bar{\beta}} \right) \ .
     \end{eqnarray}
If this equation Substituting into the Friedmann equation, we can
find the potential of the scalar field as\,\citep{Ari07}
\begin{eqnarray}
    V = 3\bar{\beta} H^2 \left[ 1-{2\over 3}
    \eta\bar{\beta}\left({\bar{\beta}_{,\phi}\over \bar{\beta}}
    + {H_{,\phi}\over H}\right)^2 \right] \ .
\end{eqnarray}
where $H=H(\phi(t))$. The equation of state find in following form
\begin{eqnarray}
    \omega_\phi &=& -1 + {4\over 3}\eta\bar{\beta}\left(\frac{H_{,\phi}}{H} +
    \frac{\bar{\beta}_{,\phi}}{\bar{\beta}} \right)^2 \nonumber\\
    &=&-1 + {1\over 3}\eta \frac{\phi^{\prime 2}}{\bar{\beta}} \ .
    \end{eqnarray}
Relations (53) and (55) have main role in following considerations.
We can see violation of the Lorentz invariance which has been
described by existence of a vector field in the action, now it has
combined in the dynamics of scalar field and equation of state by
$\bar{\beta}$ parameter. This situation allows to consider phantom
divided line crossing in LIV model. Now we should solve equations,
(53) and (55),\, to obtain dynamics of scalar field $\phi$ and the
equation of state $\omega_\phi$. Therefore, we must define the
Hubble parameter $H(\phi)$ and the vector field,\,
${\bar{\beta}}(\phi)$. In following, we choose a general case of the
Hubble parameter $H(\phi)$ and the vector field
${\bar{\beta}}(\phi)$ and then consider crossing of phantom divided
line, late time acceleration and Rip singularity scenario.

\subsection{Considering Late-Time Acceleration in LIV model}

For an acceleration universe, should be $\ddot{a}>0$. We rewrite it
to this form $H'/H > -1$. For to consider this case, we must
determine some equations. We study a general case for the vector
field and the Hubble parameter, note that these equations are a
function of $\phi$ \citep{sad08}
\begin{equation}
    H=H_0\phi^{\zeta} \ , \quad  \bar{\beta}(\phi) = m\phi^n \ ,\quad n >2 \
        \end{equation}
Following, we just study a quintessence scalar field by $\eta = 1$.
By using relation (53), we calculate
\begin{equation}
\phi \left( t \right) = \frac{1}{A}
\end{equation}
where
\begin{eqnarray}
A=\Big[H_0(t-t_0)(-4\,\zeta m+4\,\zeta mn+2\,{\zeta}^{2}m \nonumber
\end{eqnarray}
\begin{eqnarray}
~~~~-4 \,mn+2\,m{n}^{2})+\phi_0 \Big ] ^{ \left(\frac{1}{ n+\zeta-2}
\right) } \nonumber
\end{eqnarray}
and by using relation (55), we can obtain
\begin{equation}
\omega_\phi(t)=-1+\frac{4}{3}m\phi^{n-2}(t)(\zeta+n)^2.
\end{equation}
Following with relations (56) and (55,58), we find
\begin{equation}
m^2<\frac{1}{4(-1)^n\phi^{n-2}(t)(\zeta+n)^2},~~~~~~ n>2 ,
\end{equation}
This equation describe a restriction in LIV parameters for a
late time acceleration.\\
Now with using equation (56) and (57) for obtain scale factor as
\begin{equation}
a(t)=a_0(t_0)e^{\frac{2\phi_0^{\frac{\zeta}{(n+\zeta-2)}}\Big(m(n+\zeta)(n+\zeta-2)H_0(t-t_0)+\frac{1}{2}\phi_0\Big)e^{F}-\phi_0}{2\phi_0^{\frac{\zeta}{(n+\zeta-2)}}(n-2)m(n+\zeta)}}
\end{equation}
where
\begin{eqnarray}
F=\zeta
ln\Bigg(\frac{1}{e^\frac{ln(2m(n+\zeta)(n+\zeta-2)H_0(t-t_0)+\phi_0)}{n+\zeta-2}}\Bigg).
\nonumber
\end{eqnarray}
We should select a special space parameter, therefore we approach
equation (60) by a Taylor series in special space parameter
\begin{equation}
a(t)=.286504+.447663t+.237821t^2+.0352651t^3
\end{equation}
\begin{figure}[h]
\begin{center}\includegraphics{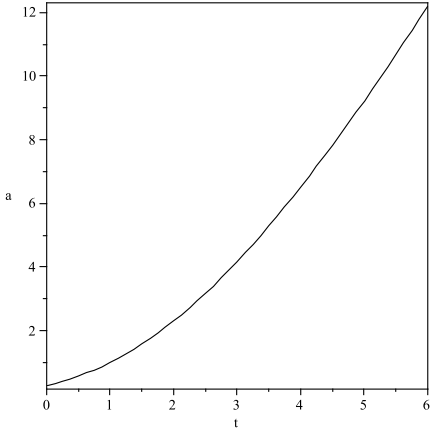} \vspace{6.2cm}
\end{center}
 \caption{\small {Variation of scale factor $a(t)$
for different values of $t$ for $n=3$ , $\zeta=-2$ and $m=-0.1$. The
values of $\zeta$ are determined by relation (59). }}
\end{figure}
We can see in figure 5 acceleration evolution in late time. Now, we
study equation of state for determine crossing phantom divided line.
By $\phi$ according to equation (57), the equation of state obtain
as
\begin{equation}
\omega_\phi(t)=-1+\frac{4}{3}m\frac{(\zeta+n)^2}{E}
\end{equation}
where \begin{eqnarray} E=[H_0(t-t_0)(-4\,\zeta m+4\,\zeta
mn+2\,{\zeta}^{2}m \nonumber
\end{eqnarray}
\begin{eqnarray}
~~~-4 \,mn+2\,m{n}^{2})+\phi_0] ^{ \left(\frac{n-2}{ n+\zeta-2}
\right) },\nonumber
\end{eqnarray}

which implicity has a dynamical behavior. This model lets us to
choose a suitable parameter space to crossing phantom divided line.
Moreover this parameter space must be compared by observational
data. We insist in this model, a scalar field and a vector field
together can explain crossing phantom divided barrier and late time
acceleration.
\begin{figure}[h]
\begin{center}\includegraphics{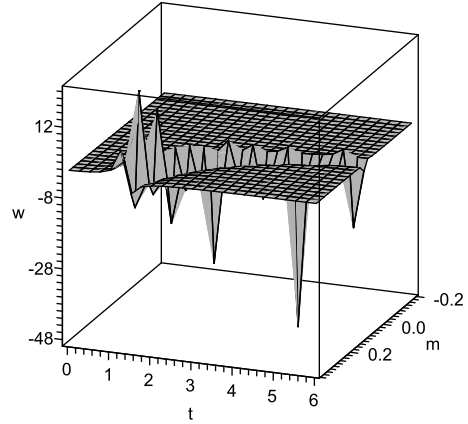} \vspace{6.2cm}
\end{center}
 \caption{\small {$\omega_\phi$ vary
for different values of the vector field by $m$ and $t$ for $n=3$
and $\zeta=-2$. Positive values of $\zeta$ does not show a phantom
divided line crossing. The values of $\zeta$  are obtained with
equation (59). }}
\end{figure}
We can see in figure 6, the crossing of phantom divided line for a
dynamical equation of state. Figure 6 perhaps explain why we are
living in an era of \, $\omega< -1$. Note that $\bar{\beta}(t)$ has
the main role of Lorentz invariance violation model. The dynamical
equation for $\bar{\beta}(t)$ has an interested mean: with a
suitable fine tuning we can obtain a Lorentz violating
cosmology correspond to observational data.\\
We know many different models that explain phantom divided line
crossing, but this model is special because it contains only a
scalar field and a Lorentz violating vector field that checks the
crossing \citep{Li08,Lib07,Ber04}. However, two notice must be
explained in this paper: First, in figures 3 and 6 we can see, there
are some sudden jerks of the equation of state. In some models that
equation of state crossing the phantom divided line, $\omega$ surge
around $-1$ (see \citep{Wei06} and references therein). These jerks
are certainly a signature of chaotic manner of equation of state.
Second, in those figures we see, crossing of the phantom divided
line appear at late-time. This means, as second cosmological
coincidence problem, requires extra fine-tuning in model
parameters.\\
For describing the physical reasons why the crossing of the phantom
divide can occur in this model, we can state by a fitting coupling
between a quintessence scalar field and other matter content may
causes to a constant ratio of the energy densities of both contexts
which is reconcilable with an accelerated expansion of the universe
or crossing of phantom dividing line. In this model, we have three
kinds of energy-momentum: 1- standard matter, 2- Scalar Field as a
nominee of Dark Energy and 3- energy-momentum content rely on for
Lorentz violating vector field. Here we take standard matter has
tiny contribution on the all energy-momentum content of the
universe. For other two energy-momentum contents, it is possible to
take the "trigger mechanism" to determined dynamical equation of
state. This means that we take scalar- vector-tensor theory
containing Lorentz invariance violation which performs like the
hybrid inflation models. In this condition, vector and scaler field
play the roles of inflaton and the "waterfall" field. Hence, it is
reasonable to  hope that one of them will eventually control to
describe inflation or accelerating phase and crossing of phantom
divided line.

\subsection{Rip Singularity}
According to Rip singularity scenario that explained in previous
section(page 5), we consider some types of Rip singularity in this
LIV model. As shown in figures (5,6), we provide suitable conditions
for considering Rip singularity scenario. Hence, correspond to
equation of state, and using equations such as energy density,
pressure and Hubble parameter, we can study some Rip singularity
solutions.\\
According our setup in this paper that selected $\zeta=-2$ and $n=3$
in anzats (56) we have
$$ H=H_0\phi^{-2} ~~,~~\beta=m\phi^{3}$$
Now with substituting above equations in energy density and pressure
relations, we can consider dynamical behaviors in these relations.
Note that, we selected for ordinary matter $\rho_{m}=0$ and
$p_{m}=0$. Now after numerical calculations (figure 7), we obtain results as following:\\
In finite time, scalae factor and energy density increases, but
pressure decreases. Therefore type II singularity (sudden) do not
predict. But type III singularity appears because $|p|$ decrease so
rapidly. In infinite time, Hubble rate tends to zero, therefore
Little Rip singularity is impossible, but pseudo-rip singularity can
occur.
\begin{figure}[h]
\begin{center}\includegraphics{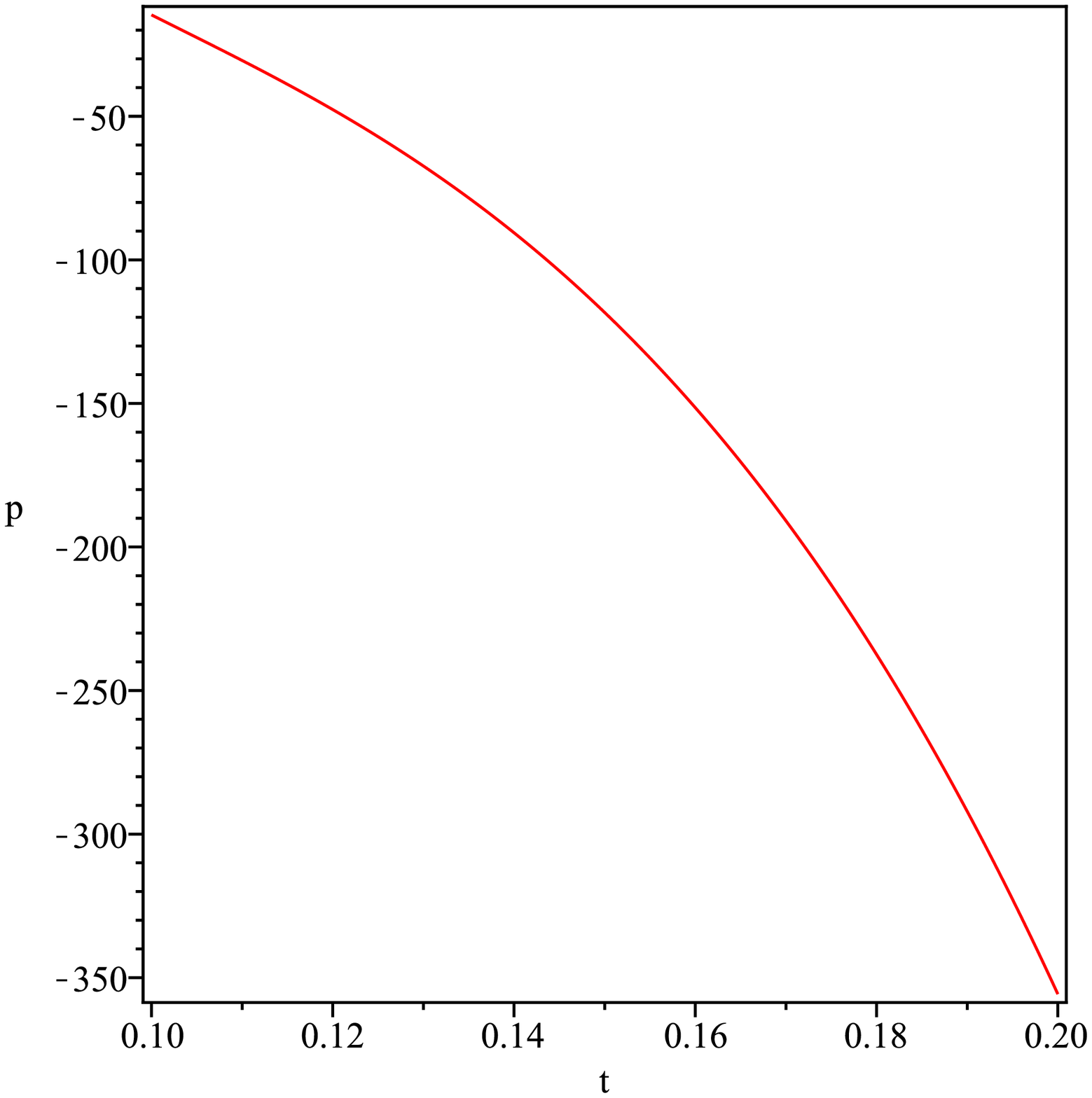} \vspace{11cm}\includegraphics{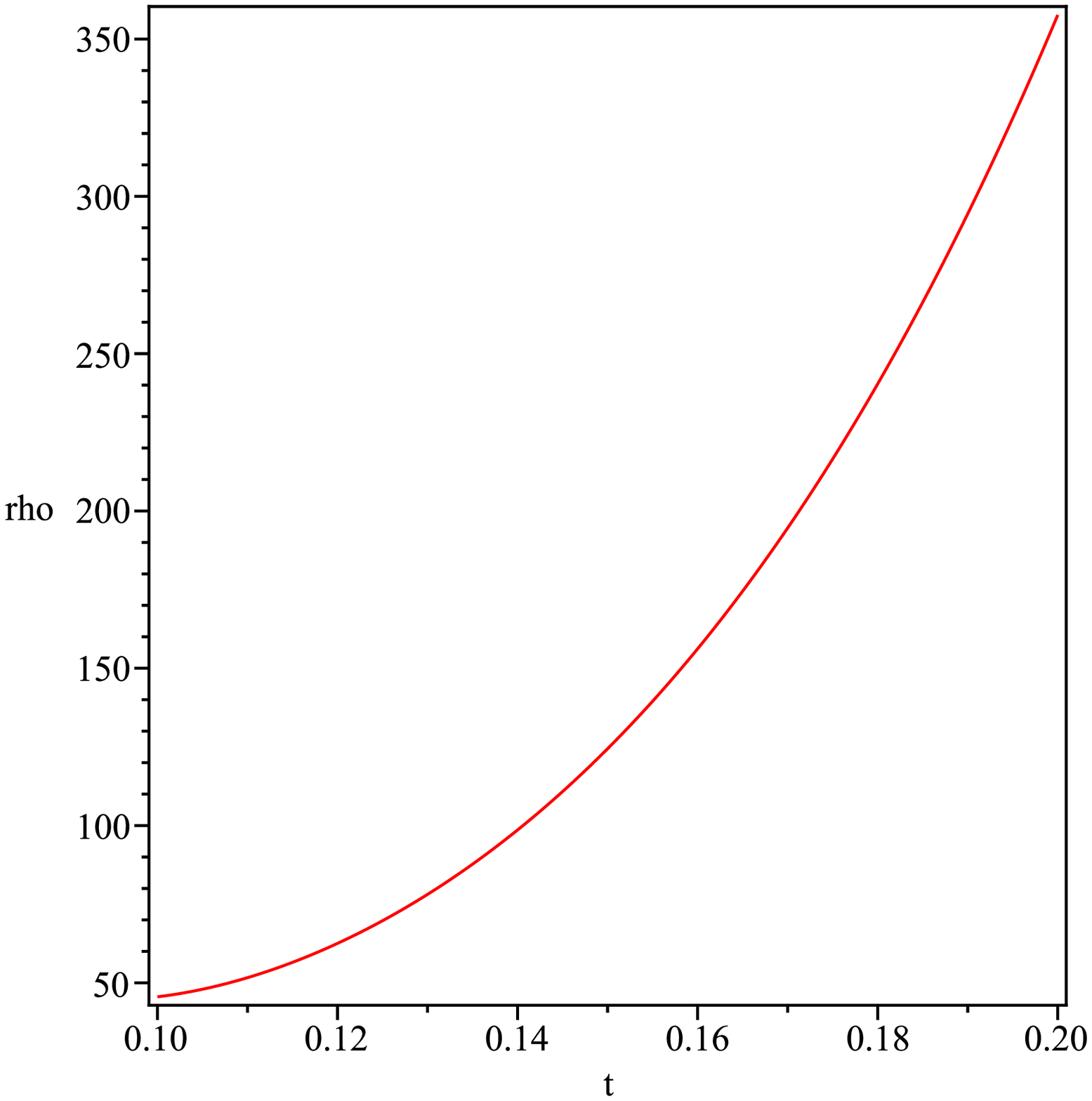}
\end{center}
\caption{\small {Variation of pressure(up) and energy density(down)
respect to time for a minimal coupling model(where $\zeta=-2$ and
$n=3$) }}
\end{figure}
We emphasize this relations determine in original setup in this
paper. This means by different selection in parameters of $\zeta$
and $m$, all dynamic of relations be change. Therefore, one can
change setup and consider other options. However, in this situation,
our work be limited.
\section{Comparison with Observational Data \&  Other Dark energy Models}
\subsection{ Analyzing SNe Data}
The parameters of the cosmological models may be consider from a
explicit comparison of their predictions with precise observational
data. Here we determined the data coming by SNe observations, the
evolution of the Hubble parameter in the brane model.

The modulus $\mu$ vs redshift $z=a_{0}/a-1$ equation corresponding
to type Ia supernovae from the Supernova Cosmology Project
\citep{Ama10,Uni2} is,
$$ \mu(z)=\mu_{0}+5\lg D_{L}(z).$$ The equation for the luminosity
distance $D_{L}(z)$ as a function of the redshift in the FRW
cosmology (FC) give as
\begin{equation}
D^{FC}_{L}=\frac{c}{H_{0}}(1+z)\int_{0}^{z} h^{-1}(z)dz,
\end{equation}
where
$h(z)=\left[\Omega_{m0}(1+z)^{3}+\Omega_{D0}F(z)\right]^{1/2}.$\\
Here, $\Omega_{m0}$ is the total fraction of matter density,
$\Omega_{D0}$ the fraction of DE energy density, and $H_{0}$ the
current Hubble parameter. The constant value $\mu_{0}$ conditional
upon the chosen Hubble parameter:
$$
\mu_{0}=42.384-5\log h,\quad h=H_{0}/100 \mbox{km/s/Mpc}
$$
The function $F(z)=\rho_{D}(z)/\rho_{D0}$ may be consider from the
continuity equation
\begin{equation}
\dot{\rho}_{D}-3\frac{\dot{a}}{a}g(\rho_{D})=0,
\end{equation}
which can be reobtain as
\begin{equation}
\int_{\rho_{D0}}^{\rho_{D}(z)}\frac{dy}{g(y)}=-3\ln(1+z).
\end{equation}
For simplicity, we ignore the contribution of radiation. For
cosmology on the brane (BC), Eq.(63) can be take as $$
D^{BC}_{L}=\frac{c}{H_{0}}(1+z)\int_{0}^{z} h^{-1}(z)[1+\delta
h^{2}(z)]^{-1/2}(1+\delta)^{1/2}d z$$ where for comfort the
parameter $\delta=\rho_{0}/2$ has been introduced. For the analysis
of the SNe data one needs to obtain the parameter $\chi^{2}$, which
is given by
\begin{equation}
\chi^{2}_{SN}=\sum_{i}\frac{\left[\mu_{obs}(z_{i})-
\mu_{th}(z_{i})\right]^{2}}{\sigma^{2}_{i}},
\end{equation}
where $\sigma_{i}$ is the corresponding $1\sigma$ error. The
parameter $\mu_{0}$ is free of the data points and, therefore, one
has to do a uniform marginalization over $\mu_{0}$. Minimization
with respect to $\mu_{0}$ may be done by simply enlarging the
$\chi^{2}_{SN}$ with respect to $\mu_{0}$,
\begin{equation}\label{chi}
\chi^{2}_{SN}=A-2\mu_{0}B+\mu_{0}^{2}C,
\end{equation}
where
$$
A=\sum_{i}\frac{\left[\mu_{obs}(z_{i})-\mu_{th}(z_{i};\mu_{0}=0)\right]^{2}}{\sigma^{2}_{i}},
$$
$$
B=\sum_{i}\frac{\mu_{obs}(z_{i})-\mu_{th}(z_{i})}{\sigma^{2}_{i}},\quad
C=\sum_{i}\frac{1}{\sigma^{2}_{i}}.
$$
The expression has a minimum for $\mu_{0}=B/C$ at
$$
\bar{\chi}_{SN}^{2}=A-B^{2}/C.
$$
One may minimize $\bar{\chi}_{SN}^{2}$ instead of
${\chi}_{SN}^{2}$.\\
with \citep{Nes05} and table 1, one obtains the 56.01\% confidence
level by $\Delta\chi^{2}=\chi^{2}-\chi^{2}_{min}<1.01$ for the
one-parameter or $1.9$ for the two-parameter models.
correspondingly, the 91.2\% confidence level is determined by
$\Delta\chi^{2}=\chi^{2}-\chi^{2}_{min}<3.78$ or $5.63$ for the one-
and two-parameter models.\\
\begin{table}
\label{Table1}
\begin{centering}
\begin{tabular}{|c|c|c|c|}
  \hline
  $z$ & $H_{obs}(z)$ & $\sigma_{H}$  \\
      & km s$^{-1}$ Mpc$^{-1}$        &  km s$^{-1}$ Mpc$^{-1}$     \\
  \hline
  0.170 & 83 & 8  \\
  1.530 & 140 & 14 \\
  1.750 & 202 & 40 \\
  \hline
\end{tabular}
\caption{Hubble parameter in contrast to
 redshift data.}
\end{centering}
\end{table}
\subsection{Dark energy Cosmology}
In this paper, we consider two models that explain dark energy in
content of the universe, but according \citep{Bam12} and references
therein, there are many different models in this subject. Therefore,
In this section for a worthy review, we briefly point out them.\\
As we know some number of well-liked dark energy models, such as the
$\Lambda$CDM model, Pseudo-Rip Little Rip and Little Rip scenarios,
the phantom and quintessence cosmology with the four types of the
singularities and non-singular universes satisfied with dark energy
have been considered.

Investigations have been shown the $\Lambda$CDM model and various
cosmological observations to endure the bounds on the late-time
acceleration of the universe. In this regards, researchers have
considered a fluid depiction of the universe in which the dark fluid
has a popular form of the Equation of state with the inhomogeneous.
They have shown that all the dark energy cosmology can be describe
by some different fluids, also determined their properties. It has
also been display that at the present stage the cosmological
evolutions of all the dark energy universes can be like to that of
the $\Lambda$CDM scenario, and hence those models are satisfied with
the cosmological observations. In other hand, they have studied the
equality of appreciate of different dark energy models, this means,
single and multiple scalar field theories could be portray. \\
However, we emphasize the role of cosmography in this review. It is
a essential tool because it lets, in doctrine, to separate among
models without a seizures but laying on restriction coming from
data. However, the goal criticism to this approach is depended to
the expansion of the Hubble parameter. In mostly, observations could
not be enlarged at any red-shift and, in some of occasions, are not
enough to track models up to early eras. Finally, the future
observational data campaigns must fix the condition elimination the
degeneration incipient at low red-shifts and let a well insight of
models.
\section{Summary}
Several evidences from observational data have shown that our
universe is in the accelerated expansion era. In this paper, we have
shown this reality in two different models: A non-minimally coupling
scalar field identified on the brane and a scalar field coupling
minimally to gravity in LIV model. According complex of dynamical
equations, we have limited our research to some special form of
non-minimal coupling and scalar field potential. Then we have
studied special form of time evolution for the scale factor and a
scalar field. In brane world model, we have determined a equation
that explain situations where an accelerating expansion of the
universe is possible. As a consequence, we have obtained which
non-minimally coupling scalar field identified on the brane is a
suitable candidate for late-time expansion of universe. In LIV
model, we have considered a new framework for crossing of phantom
divided line with equation of state, that combined a violation of
Lorentz invariance in a cosmological model. However, we have
explained with a suitable choice of parameter space, it is possible
to have phantom divided line crossing just with a Lorentz invariance
violating vector field and a single scalar field
\citep{Nes07,Vik05}. In this regard, existence of a Lorentz
invariance violating vector field prepare a setup for crossing
phantom divided line just with a minimally coupling scalar field. In
other view point, there is the possibility of a Rip singularity by
suitable tuning in the parameters of models.\\
Comparison of non-minimal model with WMAP data takes more precise
restricts on the values of non-minimal coupling. A detailed
comparison between this results and some results of previous
considerations shows that our constraints for non-minimal coupling
with exponential potential are steadfast by holographic dark energy
and also with result of warped DGP brane model\citep{Nes07}.\\

\newpage
{\bf Appendix 1:}~{\it Proof of stability}\\\\
Now we consider the stability of our model. In order to consider the
classical stability of our model, we determined the manner of the
model in the $\omega-\omega' $ plane where $\omega'$ is the
derivative of $\omega$ with the logarithm of the scale factor
(see\citep{Ari07} for a comparable analysis)
\begin{equation}
\omega'\equiv\frac{d\omega}{d\ln a}=\frac{d\omega}{dt}\frac{dt}{d
\ln a}=\frac{\dot{\omega}}{H}.
\end{equation}
The sound speed declares the phase velocity of the inhomogeneous
perturbations of the field. Hence, we take the function $c_a$ as
\begin{equation}
{c_a}^2\equiv\frac{\dot{p}}{\dot{\rho}}.
\end{equation}
If the matter is determined as a perfect fluid, this equation could
be the adiabatic sound speed of this fluid. We emphasize with scalar
fields that do not comply perfect fluid form necessarily, this
parameter is not actually a sound speed. The conservation of the
total (scaler and vector ) field given with (45). As we view, that
relation implicity depended on $\beta$ and $H$ form. Therefore, we
can take anzats (56) with substituting in (45) as
\begin{equation}
\frac{d\rho_{total}}{dt}+3(\rho_{total}+p_{total})=0
\end{equation}
Since the dust matter complies the continuity equation and the
Bianchi identity keeps legally , total energy density fulfill the
continuity equation. From that equation, we have
\begin{equation}
\dot{\rho}_{de}=-3\rho_{de}(1+\omega_{de})
\end{equation}
where $de$ means dark energy. By using equation of state
$p_{de}=\omega_{de}\rho_{de}$, we have
\begin{equation}
\dot{p}_{de}=\dot{\omega}\rho_{de}+\omega_{de}\dot{\rho}_{de}
\end{equation}
Hence, the function ${c_a}^2$ can write as
\begin{equation}
{c_a}^2=\frac{\dot{\omega}_{de}}{-3(1+\omega_{de})}+\omega_{de}
\end{equation}
In this condition, with suitable choice of  $\zeta$ and $n$ in (56),
which we take in page(10), we calculated the $\omega-\omega'$ plane
is divided into four regions defined as
$$I:~~~\omega_{de} >
-1,~~~~~\omega'>3\omega(1+\omega)~~\Rightarrow~~ {c_a}^2 > 0$$
$$II:~~~\omega_{de} >
-1,~~~~~\omega'<3\omega(1+\omega)~~\Rightarrow~~ {c_a}^2 < 0$$
$$III:~~~\omega_{de} <
-1,~~~~~\omega'>3\omega(1+\omega)~~\Rightarrow~~ {c_a}^2 < 0$$
\begin{equation}
IV:~~~\omega_{de} < -1,~~~~~\omega'<3\omega(1+\omega)~~\Rightarrow~~
{c_a}^2 > 0
\end{equation}
As one can see from these equation, the regions I and IV have the
classical stability in this model.\\\\

{\bf Appendix 2:}~{\it initial conditions for numerical calculations in both models}\\

\textbf{A} : In the brane world model\\
In the induced gravity model we should add the extra condition
$$\rho\gg |6\alpha(H^2+\frac{K}{a^2})|$$
Hence we need both high energy and weak coupling. In other hand, the
last condition and the continuity equation hints that $P = -\rho$
for $H\neq 0$. However that unlike 4D general relativity, we do not
necessarily claim $\rho =constant$. In other view point, initial
values for parameters $ \xi$ and $\beta$ given from equation
$$\xi\leq\frac{1}{16}\frac{(3\nu-1)^{2}}{2\nu(2\nu-1)}$$
and
$$\beta=\frac{3\nu-1}{2}\pm\bigg[\frac{(3\nu-1)^2}{2}-16\xi\nu(2\nu-1)\bigg]^{\frac{1}{2}}.$$
\\

\textbf{B} : In Lorentz invariance Violation Model\\
The dynamical attractor of the cosmological system has been occupied
to make the late time manners of the model uncaring to the initial
condition of the field and therefore alleviates the fine tuning
problem. In quintessence models, the dynamical system has tracking
attractor that makes the quintessence develops by tracking the
equation of state of the background cosmological fluid so as to
relieving the fine tuning problem.

Hence, in order to make viable Lorentz violation model, we need that
the coupling function $\bar{\beta}$ and the potential function V
must satisfy the condition $$\frac{\bar{\beta}
\bar{\beta}_{,\phi\phi}}{\bar{\beta}_{,\phi}^2} > 1/2$$ and
$$\frac{V V_{,\phi\phi}}{V_{,\phi}^2} +{1\over
2}{\bar{\beta}_{,\phi}/\bar{\beta}\over V_{,\phi}/V} > 1$$
respectively.


\end{document}